\documentclass[12pt]{iopart}
\def\slashchar#1{\setbox0=\hbox{$#1$}           % set a box for #1
   \dimen0=\wd0                                 % and get its size
   \setbox1=\hbox{/} \dimen1=\wd1               % get size of /
   \ifdim\dimen0>\dimen1                        % #1 is bigger
      \rlap{\hbox to \dimen0{\hfil/\hfil}}      % so center / in box
      #1                                        % and print #1
   \else                                        % / is bigger
      \rlap{\hbox to \dimen1{\hfil$#1$\hfil}}   % so center #1
      /                                         % and print /
   \fi}                                         %
\newcommand{\blank}{}
\renewcommand{\theequation}{\blank \arabic{equation}}
\newcounter{dummy}{}
\newcommand{\letters}{%
    \setcounter{dummy}{\value{equation}}
    \renewcommand{\thedummy}{\blank \arabic{dummy}}
    \renewcommand{\theequation}{\thedummy\alph{equation}}
    \refstepcounter{dummy}
    \setcounter{equation}{0}%
 }
\newcommand{\noletters}{%
    \setcounter{equation}{\value{dummy}}
    \renewcommand{\theequation}{\blank\arabic{equation}}%
  }
\newenvironment{mathletters}{\letters}{\noletters}
\newcommand{\dirac}{-i\slashchar{\nabla}}

\begin{document}
\title{Modified Spectral Boundary conditions in the Bag Model}

\author{A A Abrikosov,~jr.}

\address{ITEP, B. Cheremushkinskaya str., 25, 117 259 Moscow, Russian Federation }
\ead{persik@itep.ru}
% \date{}
% \maketitle
\begin{abstract}
We propose a reduced form of Atiah-Patodi-Singer spectral boundary
conditions for odd ($d$) dimensional spatial bag evolving in even
($d+1$) dimensional space-time. The modified boundary conditions
are manifestly chirally invariant and do not depend on time. This
allows to apply Hamiltonian approach to confined massless fermions
and study chirality effects in spatially closed volume. The
modified boundary conditions are equally suitable for chiral
fermions in Minkowski and Euclidean metric space-times.
\end{abstract}


\vspace{2pc}


\section*{Introduction}

The two principal problems of QCD are confinement and spontaneous
breaking of chiral invariance. Both phenomena take place in the
strongly interacting domain where the theory becomes
nonperturbative. Most probably they are interrelated. However,
usually they were considered separately. Up to now the spontaneous
chiral invariance breaking (SCIB) was discussed mostly in the
infinite space. It would be interesting to study specific features
of SCIB that appear due to localization of quarks in finite
volume. In order to do that one needs a chiral invariant model of
confinement.

There exists a rich family of bag models. The first was the famous
MIT bag \cite{MIT} that successfully reproduced the spectrum and
many features of hadrons. A generalization of the MIT model are
so-called chiral bags \cite{wipf/duerr,esposito/kirsten}. An
apparent drawback of these models is that the boundary conditions
are explicitly chirally noninvariant.

Attempts to save the situation led to the so called cloudy bag
model \cite{theberge} where the chiral symmetry was restored by
pion emission from the bag surface (the pion cloud). But this
model is sensitive to details of the adopted scheme of quark-pion
interaction. Thus neither of the listed models is suited to the
discussion of SCIB in finite volume.

A way to lock fermions in finite volume without spoiling the
chiral symmetry is to impose the so-called \textbf{spectral
boundary conditions} (SBC). They were first introduced by Atiah,
Patodi and Singer (APS) who investigated anomalies on manifolds
with boundaries \cite{APS}. Later these boundary conditions were
widely applied in studies of index theorems on various manifolds
\cite{euguchi/gilkey/hanson}.

Unlike the already mentioned ones the APS conditions are nonlocal.
They are defined on the boundary as a whole. This looks natural
for finite Euclidean manifolds but is inconvenient for physical
models where the time evolution takes place. The evolution
converts the spatial boundary of static physical bag into an
infinite space-time cylinder. Constraining fields on the entire
\textit{world cylinder\/} including both ``the past'' and ``the
future'' contradicts causality and complicates generalization to
Minkowski space.

In order to avoid this difficulty we propose a purely spatial
version of spectral boundary conditions. These modified conditions
do not depend on time and, therefore, are acceptable from the
physical point of view. Besides, they make possible the usual
Hamiltonian description of the system and may be used in Minkowski
space-time.

The paper has the following structure. We shall review the
classical APS boundary conditions in Section~1. In Section~2 we
shall formulate the modified spectral conditions and discuss their
properties. At the end we shall summarize the results and mention
the prospects.

\section{The APS boundary conditions and their physics}

\subsection{Conventions}

We will start from the traditional form of SBC. First we will
introduce coordinates, Dirac matrices and the gauge that allow to
most clearly define the spectral boundary conditions. For
simplicity we will consider the 4-dimensional case. The
generalization to higher even dimensions is straightforward.

Let us consider massless fermions interacting with gauge field
$\hat{A}$ in a closed Euclidean domain $B_4$.  We choose the
curvilinear coordinates so that near the boundary
 $\partial B_4$ the first coordinate $\xi$ points along the outward
normal while the three others, $q^i$, parametrize $\partial B_4$
itself. The origin $\xi=0$ lies on $\partial B_4$. For simplicity
we shall assume that near the surface the metric
 $g_{\alpha \beta}$ depends only on $q$ so that
\begin{equation}\label{g-APS}
  ds^2 =  d \xi^2 + g_{ik}(q)\, dq^i\, dq^k .
\end{equation}
Moreover, we choose the gauge so that on the boundary
 $\hat{A}_\xi = 0$.

Now we must fix the Dirac matrix $\gamma^\xi$. Let $I$ be the
$2\times 2$ unity matrix. Then
\begin{equation}\label{gamma-APS}
  \gamma^\xi = \left( \begin{array}{cc}
  0         & iI \\
  -iI & 0
\end{array} \right);
\qquad \qquad
  \gamma^q = \left( \begin{array}{cc}
  0         & \sigma^q \\
  \sigma^q  & 0
\end{array} \right).
\end{equation}
Matrices $\sigma^q$ are the ordinary Pauli $\sigma$-matrices. With
these definitions the Dirac operator of massless fermions on the
surface takes the form,
\begin{equation}\label{nabla-APS}
    \left.
        -i \slashchar{\nabla}
    \right|_{\partial B_4} =
         -i \gamma^\alpha \nabla_\alpha =
 \left(\begin{array}{cc}
  0                 & \hat{M}  \\
  \hat{M}^\dagger   & 0
\end{array} \right) =
 \left(\begin{array}{cc}
  0                 & I \, \partial_\xi - i \hat{\nabla}  \\
   - I \, \partial_\xi - i \hat{\nabla} & 0
\end{array} \right) ,
\end{equation}
where $\hat{\nabla} = \sigma^q\, \nabla_q$ is the convolution of
covariant gradient along the boundary $\nabla_q$ with
$\sigma$-matrices. Note that Hermitian conjugated operators
$\hat{M}$ and $\hat{M}^\dagger$ differ only by the sign of
$\partial_\xi$-derivative.

Further on we shall call the covariant derivative
 $-i \hat{\nabla}$ on the boundary the \textbf{boundary operator.} It
is a linear differential operator acting on 2-spinors. It is
Hermitian and includes tangential gauge field $\hat{A}_q$ and
the spin connection which arises from the curvature of $\partial
B_4$.

The massless Dirac operator anticommutes with $\gamma^5$-matrix:
\begin{equation}\label{chirality}
  \left \{-i \slashchar{\nabla},\, \gamma^5 \right\} = 0,
\qquad
        \gamma^5 = \left(
\begin{array}{rr}
  I & 0 \\
  0 & -I
\end{array} \right),
\end{equation}
and gauge  interactions do not change helicity of massless quarks.
This property is  called chiral invariance. In order to preserve
it in finite space one needs chirally invariant boundary
conditions.

\subsection{The APS boundary conditions} \label{APS}
\subsubsection{The definition}

Atiah, Patodi and Singer investigated spectra of Dirac operator on
manifolds with boundaries. If we separate upper and lower (left
and right) components of 4-spinors the corresponding eigenvalue
equation for $-i\slashchar{\nabla}$ will take the form
\begin{equation}\label{eigenvalue}
 -i\slashchar{\nabla}\, \psi_\Lambda =
 -i\slashchar{\nabla} \left(
\begin{array}{c}
u_\Lambda  \\ v_\Lambda
\end{array}\right) =
 \Lambda \left(
\begin{array}{c}
u_\Lambda  \\ v_\Lambda
\end{array}\right ) =
 \Lambda\, \psi_\Lambda.
\end{equation}
The next step is to Fourier-expand $u$ and $v$ near the boundary.
Let 2-spinors $e_\lambda (q)$ be eigenfunctions of the boundary
operator $-i\hat{\nabla}$:
\begin{equation}\label{e-lambda}
  -i\hat{\nabla}\, e_\lambda (q) = \lambda\, e_\lambda (q).
\end{equation}
Note that the form of this equation and the eigenfunctions
$e_\lambda (q)$ depend on gauge. It is here that the gauge
condition
 $\hat{A}_\xi (0,\, q) = 0$ becomes important.

The operator $-i\hat{\nabla}$ is Hermitian so $\lambda$'s are
real. The functions $e_\lambda$ form an orthogonal basis on
 $\partial B_4$. In principle $-i\hat{\nabla}$ may have zero-modes
but sphere and convex manifolds are not the case.

In the vicinity of the boundary spinors $u_\Lambda$ and
$v_\Lambda$ may be expanded in series in $e_\lambda$:
\begin{mathletters}\label{uv-exp}
\begin{eqnarray}
  u_\Lambda (\xi,\, q) & = &
    \sum_\lambda f_\Lambda^\lambda (\xi)\, e_\lambda (q),
\quad
    f_\Lambda^\lambda (\xi)
        = \int_{\partial B_4}
        e_\lambda^\dagger (q)\, u_\Lambda (\xi,\, q)\,
        \sqrt g\, d^3 q;
\label{uv-exp/a} \\
  v_\Lambda (\xi,\, q) & = &
    \sum_\lambda g_\Lambda^\lambda (\xi)\, e_\lambda (q),
\quad
    g_\Lambda^\lambda (\xi)
        = \int_{\partial B_4}
        e_\lambda^\dagger (q)\, v_\Lambda (\xi,\, q)\,
        \sqrt g\, d^3 q;
\label{uv-exp/b}
\end{eqnarray}
\end{mathletters}
where $g=\det ||g_{ik}||$ is the determinant of metric on the
boundary.

The spectral boundary conditions state that on the boundary,
\emph{i.~e.\/} at $\xi =0$
\begin{mathletters}\label{SBC}
\begin{eqnarray}
  \left.
    f_\Lambda^\lambda
  \right|_{\partial B_4} & = & 0
    \qquad \mathrm{for} \qquad \lambda > 0;  \label{SBC/a} \\
  \left.
    g_\Lambda^\lambda
  \right |_{\partial B_4} & = & 0
    \qquad \mathrm{for} \qquad \lambda < 0.  \label{SBC/b}
\end{eqnarray}
\end{mathletters}
Another way to say this is to introduce integral projectors
$\mathcal{P}^+$ and $\mathcal{P^-}$ onto boundary modes with
positive and negative $\lambda$:
\begin{equation}\label{Projectors}
  \mathcal{P}^+ (q,\, q') =
    \sum_{\lambda > 0} e_\lambda (q) \, e_\lambda^\dagger (q');
\qquad
  \mathcal{P}^- (q,\, q') =
    \sum_{\lambda < 0} e_\lambda (q) \, e_\lambda^\dagger (q').
\end{equation}
Let $\mathcal{I}$ be the unity operator on the function space
spanned by $e_\lambda$. Then, obviously,
\begin{equation}\label{Ppls/Pmns}
    \mathcal{P^+ + P^- = I .}
\end{equation}
If we join two-dimensional projectors $\mathcal{P}^+$ and
$\mathcal{P^-}$ into $4\times 4$ matrix $\mathcal{P}$ the spectral
boundary condition  for 4-spinor $\psi$ will look as follows:
\begin{equation}\label{P-APS}
    \left.
        \mathcal{P}\, \psi
    \right|_{\partial B_4} =
    \left.
    \left(
        \begin{array}{cc}
          \mathcal{P}^+ & 0 \\
          0 & \mathcal{P}^- \\
        \end{array}
    \right)
    \left(
        \begin{array}{c}
          u \\
          v \\
        \end{array}
    \right)
    \right|_{\partial B_4} = 0.
\end{equation}
The projector $\mathcal{P}$ commutes with matrix $\gamma^5$:
\begin{equation}\label{[P,gamma5]}
    \left[
        \mathcal{P},\, \gamma^5
    \right] = 0.
\end{equation}
Therefore boundary condition (\ref{P-APS}) by construction
respects chiral invariance.

\subsubsection{The physics}

Now we shall prove that the spectral boundary conditions are
acceptable and explain their physical meaning. Namely, we shall
show that SBC provide Hermicity of the Dirac operator and
conservation of fermions in the bag. After that we will explain
the origin of requirements (\ref{SBC}).

First let us prove that Dirac operator is Hermitian. As usually,
we integrate by parts the expression
\begin{equation}\label{Dirac=Herm}
    \int_{B_4} dV\, f^\dagger\, (\dirac g) =
    \int_{B_4} dV\, (\dirac f)^\dagger\, g +
    \oint_{\partial B_4} dS\, f^\dagger\, (-i \gamma^\xi) g.
\end{equation}
Now we need to show that if $f$ and $g$ satisfy (\ref{SBC}) then
the last term vanishes.

Conditions (\ref{SBC}) mean that on the boundary 4-spinors $f$ and
$g$ may be written as:
    $f = \left(
        \begin{array}{c}
          f^- \\
          f^+ \\
        \end{array}
        \right)$
and
    $g = \left(
        \begin{array}{c}
          g^- \\
          g^+ \\
        \end{array}
        \right)$,
where $f^\pm$ and $g^\pm$ include only components with positive
and negative $\lambda$ respectively, see (\ref{uv-exp}). Rewriting
the boundary term in (\ref{Dirac=Herm}) we get
\begin{equation}\label{oint=0}
    \oint_{\partial B_4} dS\, f^\dagger\, (-i \gamma^\xi) g =
    \oint_{\partial B_4} dS\,
    \left[
        (f^-)^\dagger\, g^+ - (f^+)^\dagger\, g^-
    \right] = 0,
\end{equation}
due to the orthogonality of eigenfunctions of the boundary
operator. Thus the APS boundary conditions indeed ensure the
Hermicity of Dirac operator.

In addition, relation (\ref{oint=0}) guarantees conservation of
fermions in the bag. Indeed, for $f = g$ the LHS is nothing but
the net fermionic current through the boundary,
\begin{equation}\label{Fconserv}
    \oint_{\partial B_4} dS\, j^\xi =
    - i\oint_{\partial B_4} dS\, f^\dagger \gamma^\xi f = 0.
\end{equation}
Therefore the number of fermions is conserved and particles in the
spectral bag are confined.

In order to understand the physics of SBC let us rewrite the
eigenvalue condition (\ref{eigenvalue}) near the boundary in terms
of components.
\begin{mathletters}\label{fg-eqns}
\begin{eqnarray}
  (\partial_\xi + \lambda)\, g_\Lambda^\lambda (\xi) & = &
    \Lambda\, f_\Lambda^\lambda (\xi);
        \label{fg-eqns/a} \\
  - (\partial_\xi - \lambda)\, f_\Lambda^\lambda (\xi) & = &
    \Lambda\, g_\Lambda^\lambda (\xi).
        \label{fg-eqns/b}
\end{eqnarray}
\end{mathletters}
Depending on the sign of $\lambda$ these relations reduce on the
boundary either to
\begin{mathletters}\label{logD}
\begin{equation}\label{logD/a}
  \left. \frac{\partial_\xi g_\Lambda^\lambda }{g_\Lambda^\lambda }
    \right|_{\xi=0}
    = -\lambda <0 , \qquad f_\Lambda^\lambda (0) = 0
        \quad \mathrm{at} \quad \lambda > 0;
\end{equation}
or to
\begin{equation}\label{logD/b}
  \left. \frac{\partial_\xi f_\Lambda^\lambda }{f_\Lambda^\lambda }
    \right|_{\xi=0}
    = \hphantom{-} \lambda <0 , \qquad g_\Lambda^\lambda (0) = 0
        \quad \mathrm{at} \quad \lambda < 0.
\end{equation}
\end{mathletters}
Thus both components either vanish on the boundary or have a
negative logarithmic derivative along the normal.

This requirement has a simple physical interpretation. Suppose
that out of the bag the metric and the gauge field remain the same
as on the boundary. Then we can continue the functions $f$ and $g$
outside the bag to $\xi = \infty$.  Some of the functions will be
zero, $f^+ = g^- = 0$ at $\xi > 0$, and the rest will be falling
square integrable exponents
  $f^\lambda,\, g^\lambda \propto \exp -|\lambda| \xi$ similar to
wave functions of particles locked in a potential well. The only
difference is that now the depth of the well depends on $\lambda$
and is adjusted for each mode specially. We may conclude that the
spectral boundary conditions claim that wave functions in the bag
must have square integrable continuation to infinity.

\section{The SBC for physical bags}

\subsection{The truncated SBC}

Now let us turn to fermions confined in a 3-dimensional spatial
bag $B_3$ that evolves in Euclidean time and sweeps the infinite
space-time cylinder $B_3\otimes R$. We will call the first three
coordinates ``space'' and the fourth one ``time''. The boundary
operator consists of spatial and temporal parts:
\begin{equation}\label{BOfull}
    -i \hat{\nabla}_{\partial B_3 \otimes R} =
    -i \hat{\nabla}_{\partial B_3} - i \sigma^z \partial_4 .
\end{equation}
We will call the spatial part
 $-i \hat{\nabla}_{\partial B_3}$ the
\textbf{truncated boundary operator}. Let its eigenfunctions be
$e^\pm_\lambda$:
\begin{equation}\label{e-lambda-3d}
  -i\hat{\nabla}_{\partial B_3}\, e^\pm_\lambda (q) =
  \pm\lambda\, e^\pm_\lambda (q),
\qquad
    \lambda > 0.
\end{equation}

Wave functions on the space-time boundary $\partial B_3\otimes R$
can be expanded in $e^\pm_\lambda$ and longitudinal (temporal)
plane waves:
\begin{mathletters} \label{Fourier+/-}
\begin{eqnarray}
  u_\Lambda  & = &
    \sum_{\lambda >0}
    \int \frac{dk}{2\pi}\,e^{ikt}\,
    \left[
        f^{+\lambda,\, k}_\Lambda \, e^+_\lambda +
        f^{-\lambda,\, k}_\Lambda \, e^-_\lambda
    \right] ;
\label{Fourier+/-a}\\
  v_\Lambda  & = &
    \sum_{\lambda >0}
    \int \frac{dk}{2\pi}\,e^{ikt}\,
    \left[
        g^{+\lambda,\, k}_\Lambda \, e^+_\lambda +
        g^{-\lambda,\, k}_\Lambda \, e^-_\lambda
    \right] .
\label{Fourier+/-b}
\end{eqnarray}
\end{mathletters}

The truncated operator $-i \hat{\nabla}_{\partial B_3}$
anticommutes with $\sigma^z$. Therefore $\sigma^z$ changes the
sign of $e$-eigenvalues. A possible choice of eigenvectors is (see
\cite{Leipzig01,abrikosov} for the sphere)
\begin{equation}\label{e+/e-}
  e^\pm_\lambda =
    \pm i \sigma^z \, e^\mp_\lambda .
\end{equation}
Thus the last term in (\ref{BOfull}) mixes positive and negative
spatial harmonics.

In classical approach this would mean that SBC should be written
in terms of $k$-dependent eigenfunctions of the full boundary
operator (\ref{BOfull}) which look rather complicated. Moreover,
extending boundary  conditions onto the entire interval
  $- \infty < t < \infty$ makes them ``future-sensitive'' and
violates causality. Therefore we propose to consider the
$k$-independent \textbf{truncated APS constraints}:
\begin{mathletters}\label{SBC3+1}
\begin{eqnarray}
    \left.
        f^{+\lambda,\, k}_\Lambda
    \right|_{\partial B_3} & = & 0 ;
\label{SBC3+1/a}
    \\
    \left.
        g^{-\lambda,\, k}_\Lambda
    \right|_{\partial B_3} & = & 0 .
\label{SBC3+1/b}
\end{eqnarray}
\end{mathletters}
These conditions do not depend on time and allow Hamiltonian
treatment of the system. Besides, they may be applied both in
Euclidean and Minkowski spaces. Now let us show that they are
acceptable.

\subsection{Consistency}

We are going to prove that the truncated form of SBC fulfills the
necessary conditions. Namely, they are chirally invariant, the
Dirac operator is Hermitian, the fermion number is conserved and,
after all, wave functions may be continued out of the bag to
spatial infinity.

The proof of the first three points literally follows the
4-dimensional case. Everything that concerns formulae
(\ref{Projectors}--\ref{Fconserv}) remains true for truncated
($_T$) 3-dimensional SBC (\ref{SBC3+1}). One may define on
$\partial B_3$ projectors,
\begin{equation}\label{3d-Projectors}
  \mathcal{P}^\pm_T (q,\, q') =
    \sum_{\lambda > 0} e^\pm_\lambda (q) \,
    \left[
        e^\pm_\lambda (q')
    \right]^\dagger.
\end{equation}
Then the truncated boundary conditions may be written in the
manifestly $\gamma^5$-invariant form,
\begin{equation}\label{3d-P-APS}
    \left.
        \mathcal{P}_T\, \psi
    \right|_{\partial B_3} =
    \left.
    \left(
        \begin{array}{cc}
          \mathcal{P}_T^+ & 0 \\
          0 & \mathcal{P}_T^- \\
        \end{array}
    \right)
    \left(
        \begin{array}{c}
          u \\
          v \\
        \end{array}
    \right)
    \right|_{\partial B_3} = 0.
\end{equation}

Hermicity of Dirac operator and conservation of fermions are
proven in the same way as before, see
(\ref{Dirac=Herm}--\ref{Fconserv}), so we skip the formulae.

The last point is more delicate. We already mentioned that the
$\sigma^z$-piece in (\ref{BOfull}) mixes positive and negative
harmonics. Therefore they must be analysed together and instead of
two eigenvalue equations (\ref{fg-eqns}) we get four ($\xi$ is the
spatial normal to the boundary):
\begin{mathletters}\label{RHS3+1}
\begin{eqnarray}
      (\partial_\xi
      + \lambda)\,
      g_\Lambda^{+\lambda,\, k}\,
      & = &
    \Lambda\,  f_\Lambda^{+\lambda,\, k}
    + ik\,  g_\Lambda^{-\lambda,\, k};
\label{RHS3+1/a}
\\
    - (\partial_\xi
     - \lambda)\, f_\Lambda^{+\lambda,\, k}
     & = &
     \Lambda\, g_\Lambda^{+\lambda,\, k}
    + ik\, f_\Lambda^{-\lambda,\, k} :
\label{RHS3+1/b}
\\
      (\partial_\xi
      - \lambda)\,
      g_\Lambda^{-\lambda,\, k}\,
      & = &
    \Lambda\,  f_\Lambda^{-\lambda,\, k}
    - ik\,  g_\Lambda^{+\lambda,\, k};
\label{RHS3+1/c}
\\
    - (\partial_\xi
     + \lambda)\, f_\Lambda^{-\lambda,\, k}
     & = &
     \Lambda\, g_\Lambda^{-\lambda,\, k}
    - ik\, f_\Lambda^{+\lambda,\, k} .
\label{RHS3+1/d}
\end{eqnarray}
\end{mathletters}
The new feature with respect to (\ref{fg-eqns}) are $ik$-terms
that appear due to the mixing.  However one may notice that the
terms in the RHS of (\ref{RHS3+1}) come in pairs $f^+$, $g^-$ and
$f^-$, $g^+$. Therefore according to conditions (\ref{SBC3+1}) the
RHS of equations (\ref{RHS3+1/a}, \ref{RHS3+1/d}) still vanish on
the boundary. Thus the behaviour of $g^+$ and $f^-$ on the
boundary is governed by the homogeneous equations and
\begin{equation}\label{logDf,g}
    \left.
        \frac{\partial_\xi f_\Lambda^{-\lambda,\, k} }%
        {f_\Lambda^{-\lambda,\, k}}
    \right|_{\xi=0} =
    \left.
        \frac{\partial_\xi g_\Lambda^{+\lambda,\, k} }%
        {g_\Lambda^{+\lambda,\, k}}
    \right|_{\xi=0} = -\lambda < 0.
\end{equation}

Hence despite the presence of extra pieces the nonvanishing
components $g^+$ and $f^-$ have negative logarithmic derivatives.
This means that solutions of eigenvalue equations may be continued
from the world cylinder swept by evolving bag to spatial infinity
in an integrable way. Thus the last of the requirements is
fulfilled. This completes the proof of acceptability of the
truncated SBC.

\section*{Conclusion}

The truncated version of APS boundary conditions offers a number
of possibilities. It allows to formulate a chirally invariant bag
model and to address chiral properties of fermionic field in the
closed volume. The constraints do not depend on time so one may
write down the Hamiltonian and study the energy (and mass)
spectrum of the system. Another advantage is that the modified SBC
may be used both in Euclidean and Minkowski spaces.

A new feature that SBC may bring to bag physics is their
nonlocality. The usually employed local boundary conditions, see
\cite{MIT,wipf/duerr,esposito/kirsten}, correspond to the thin
wall approximation. The spectral conditions refer to the boundary
as a whole. Therefore, in a sense, hadrons are also treated as a
whole which complies with modern concepts. It would be interesting
to investigate hadronic spectra in chiral invariant bags and find
out if the model is realistic and what it is missing.

Another question is more mathematical. Chiral symmetry is specific
for fermions in even-dimensional spaces. Hence the spectral
boundary conditions were always discussed in even dimensions. The
truncated SBC are formulated in the odd-dimensional space that
remains after discarding the time. This might have interesting
consequences. For example, the boundary of odd-dimensional bag is
an even-dimensional manyfold and the truncated boundary operator
possesses a sort of internal chirality. It would be interesting to
study consequences of this hidden symmetry.

In conclusion I would like to express my gratitude to Professor
A.~Wipf for fruitful discussions. I thank the Organizers for
financial support of my participation in the Conference. The work
was partially supported by RFBR grants 03--02--16209 and
\mbox{05--02--17464}.

%----------- References ----------------------------------


\begin{thebibliography}{99}
%\def\journal#1#2#3#4{{#1},\ {#3},\ {vol.~#2},\ {p.~#4}}
%\def\journal#1#2#3#4{{#1}\ #2\ (#3)\ #4}
\def\journal#1#2#3#4{{#1}\textbf{#2}\ (#3)\ #4}
%
%     Specific journals:
\def\NPB{\journal{Nucl.\ Phys.\ \textbf{B}}}
\def\PRC{\journal{Phys.\ Rep.\ \textbf{C}~}}
\def\PRD{\journal{Phys.\ Rev.\ \textbf{D}}}
\def\MPCPS{\journal{Math.\ Proc.\ Camb.\ Phil.\ Soc.\ }}
\def\IJMPA{\journal{Int.\ Journ.\ of\ Mod.\ Phys.\ \textbf{A}}}
%===================================================
    \bibitem{MIT} A.~Chodos, R.~L.~Jaffe, C.~B.~Thorn, and
        V.~F.~Weisskopf, \PRD{9}{1974}{3471}.
    \bibitem{wipf/duerr} S.~Duerr, A.~Wipf, \NPB{443}{1995}{201}.
    \bibitem{esposito/kirsten} G.~Esposito, K.~Kirsten,
        \PRD{66}{2002}{085014}.
    \bibitem{theberge} S.~Th\'eberge, A.~W.~Thomas, G.~A.~Miller,
    \PRD{22}{1980}{2838}; \PRD{23}{1981}{2106(E)}.
    \bibitem{APS} M.~F.~Atiah, V.~K.~Patodi, I.~M.~Singer,
        \MPCPS{77}{1975}{43}.
    \bibitem{euguchi/gilkey/hanson} T.~Eguchi, P.~B.~Gilkey,
        A.~J.~Hanson, \PRC{66}{1980}{213}.
    \bibitem{Leipzig01}A.~A.~Abrikosov, \IJMPA{17}{2002}{885}.
    \bibitem{abrikosov} A.~A.~Abrikosov, hep-th/0212134.
\end{thebibliography}
\end{document}